\documentclass[11pt]{article}
\usepackage{mymoriond,epsfig}

\def\be{\begin{equation}}
\def\ee{\end{equation}}
\def\bea{\begin{eqnarray}}
\def\eea{\end{eqnarray}}

\long\def\comment#1{}

\def\la{\hbox{ \raise.35ex\rlap{$<$}\lower.6ex\hbox{$\sim$}\ }}
\def\ga{\hbox{ \raise.35ex\rlap{$>$}\lower.6ex\hbox{$\sim$}\ }}

\def\W2{{\cal W}}

\def\la{\bigl\langle}

\def\be{\begin{equation}}
\def\bea{\begin{eqnarray}}
\def\ee{\end{equation}}  
\def\eea{\end{eqnarray}}

\begin{document}
\vspace*{4cm}
\title{NON--VACUUM INITIAL STATES FOR THE COSMOLOGICAL PERTURBATIONS
AND THE BACK--REACTION PROBLEM OF INFLATION}
\author{Alejandro Gangui$^{1,2}$, J\'er\^ome Martin$^3$, and Mairi Sakellariadou$^{4}$}
\address{$^1$ Instituto de Astronom\'{\i}a y F\'{\i}sica del Espacio,
              Ciudad Universitaria, 1428 Buenos Aires, Argentina   \\
         $^2$ Dept. de F\'{\i}sica, Universidad de Buenos Aires, 
              Ciudad Universitaria -- Pab. 1, 1428 Bs As, Argentina \\ 
         $^3$ IAP -- Institut d'Astrophysique
              de Paris, 98bis boulevard Arago, 75014 Paris, France  \\
         $^4$ Department of Astrophysics, Astronomy, and Mechanics,
              University of Athens, Zografos, Hellas }

\maketitle\abstracts{ In the framework of Inflationary theory, the
assumption that the quantum state of the perturbations is a non-vacuum
state leads to a difficulty: non-vacuum initial states imply, in
general, a large energy density of inflaton field quanta, not of a
cosmological term type, that could prevent the inflationary phase. In
this short note, we discuss in detail why this is so, keeping an eye
on possible non-Gaussian features due to considering generic
non-vacuum initial states.}

\noindent {\bf 1. Introduction:} 
The inflationary scenario is today the most appealing candidate for
describing the early universe. It makes four key predictions: (i) the
curvature of the space-like sections vanishes, i.e. the total energy
density, relative to the critical density, is $\Omega _0=1$, (ii) the
power spectrum of density fluctuations is almost scale invariant,
i.e. its spectral index is $n_{_{\rm S}}\simeq 1$, (iii) there is a
background of primordial gravitational waves (also scale invariant),
and (iv) the statistical properties of the cosmic microwave background
(CMB) are Gaussian.

Gaussianity for the CMB can be directly traced back to the standard
lore that the quantum fluctuations of the inflaton field are placed in
the vacuum state. Thus, one might want to test
the robustness of this prediction in the case where the vacuum state
assumption is relaxed~\cite{MRS}.  However, assuming
that the quantum state of the perturbations is a non-vacuum state
immediately leads to the problem that a large energy density of
inflaton field quanta is produced~\cite{ll}. This results in a back-reaction
problem that could upset the inflationary phase.

What we will see below is that one cannot directly conclude that this
would prevent inflation from occurring altogether because, without a
detailed calculation, it is difficult to guess what the back-reaction
effect on the background would be. Such a detailed calculation is in
principle possible by means of the formalism developed in
Ref.~\cite{Abramo}. To our knowledge, such a computation has never
been performed. The calculation of second order effects is clearly a
complicated issue and is still the subject of discussions in the
literature, see~\cite{Unruh} for example. Moreover there exist
situations where it can be avoided, and this is in fact the case if
the number of e-folds is not too large, as we explain below.

There exist other reasons to study non-vacuum initial
states~\cite{GMS}. In particular, one could imagine that the inflaton
field emerges from the trans-Planckian regime in a non-vacuum
state. Detecting somehow the presence of these non-vacuum states would
then be a signature of non standard physics and it seems interesting
to consider this issue.  From the theoretical point of view the
simplest way to generalize the vacuum initial state, which contains no
privileged scale, is to consider an initial state with a {\em
built-in} characteristic scale, $k_{\rm b}$ ~\cite{MRS}.
In what follows we discuss in detail the argument developed in
Ref.~\cite{ll} regarding the back-reaction problem. We show that any
theory with a non-vacuum initial state has to face this
issue. However, we also argue that it is not clear at all whether
inflation will be prevented in this context.

\begin{figure}[t]
\centering
\epsfig{figure=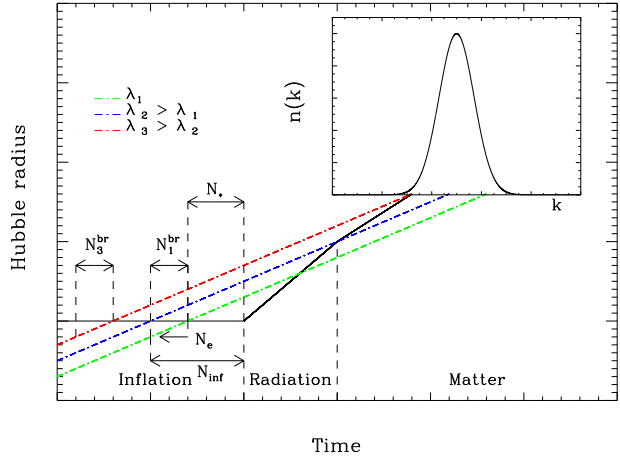, width=10cm}
\vspace*{-0.5cm}
\caption{Sketch of the evolution of the physical size in
an inflationary universe where the cosmological perturbations
are placed in a non-vacuum state characterized by the distribution $n(k)$.}
\label{breac}
\vspace*{-0.5cm}
\end{figure}

\noindent {\bf 2. Initial state for the cosmological perturbations
and the back-reaction problem:} Let us discuss the relevance of non-vacuum initial states
for cosmological quantum perturbations. The argument of Ref.~\cite{ll}
is based on the calculation of the energy density of the perturbed
inflaton scalar field in a given non-vacuum initial state. Since the
perturbed inflaton and the Bardeen potential are linked through the
Einstein equations, it is clear that they should be placed in the same
quantum state. Let us consider a quantum scalar field living in
a (spatially flat) Friedmann--Lema\^{\i}tre--Roberston--Walker
background. The expression of the corresponding operator for the
perturbations reads
\begin{equation}
\varphi (\eta ,{\bf x})=\frac{1}{a(\eta )}\frac{1}{(2\pi
)^{3/2}}
\int {\rm d}^3{\bf k}\frac{1}{\sqrt{2k}}
\biggl[\mu _k(\eta )c_{\bf k}(\eta _{\rm i})e^{i{\bf k}\cdot {\bf x}}
+\mu _k^*(\eta )c_{\bf k}^{\dagger }(\eta _{\rm i}) e^{-i{\bf k}
\cdot {\bf x}}\biggr],
\end{equation}             
where $c_{\bf k}(\eta _{\rm i})$ and $c_{\bf k}^{\dagger }(\eta _{\rm
i})$ are the annihilation and creation operators (respectively)
satisfying the commutation relation $[c_{\bf k},c_{\bf p}^{\dagger }]=
\delta ({\bf k}-{\bf p})$, and where $a(\eta )$ is the scale factor
depending on conformal time $\eta$. The equation of motion for the
mode function $\mu _k(\eta )$ can be written as
$\mu _k''+[k^2-({a''}/{a})]\mu _k=0$,
where ``primes'' stand for derivatives with respect to conformal time.
The above is the characteristic equation of a parametric oscillator
whose time-dependent frequency depends on the scale factor and its
derivative. The energy density and pressure for a scalar field are
given by 
\begin{equation}
\rho = \frac{1}{2a^2}\varphi '^2+V(\varphi )+\frac{1}{2a^2}\delta
^{ij}\partial _i\varphi \partial _j \varphi ,
\quad
p = \frac{1}{2a^2}\varphi '^2-V(\varphi )-\frac{1}{6a^2}\delta
^{ij}\partial _i\varphi \partial _j \varphi .
\end{equation}    
We can now compute the energy and pressure in a state characterized
by a distribution $n(k)$ (giving the number $n$ of quanta with comoving
wave-number $k$) for a free (i.e. $V=0$) field. Let us denote such a
state by $\vert n(k)\rangle $. Using some simple algebra 
and restricting ourselves to the high-frequency regime, 
$\mu _k\simeq \exp[-ik (\eta -\eta _{\rm i})]$, where $\eta _{\rm i}$
is some given initial conformal time, and subtracting the standard
vacuum contribution, it is easy to find the expressions for the
density and pressure in the $\vert n(k) \rangle$ state~\cite{GMS}
\begin{equation}
\langle n(k) \vert \rho \vert n(k) \rangle =
\frac{1}{2\pi ^2 a^4}\int _0^{+\infty } \frac{{\rm d}k}{k}
k^4n(k), \quad
\langle n(k) \vert p \vert n(k) \rangle =
\frac{1}{2\pi ^2 a^4}\frac{1}{3}\int _0^{+\infty } \frac{{\rm
d}k}{k}k^4n(k).
\end{equation}      
For a well-behaved distribution function $n(k)$ this result is finite.
Moreover, the perturbed inflaton (scalar) particles behave as
radiation, as clearly indicated by the equation of state $p=(1/3)\rho$
and as could have been guessed from the beginning since the
scalar field studied is free. To go further, we need to specify the
function $n(k)$. If we assume that the distribution $n(k)$ is peaked
around a value $k_{\rm b}$, it can be approximated by a constant
distribution of n quanta, with $n(k_{\rm b})\simeq n$, in the interval
$[k_{\rm b}-\Delta k,k_{\rm b}+\Delta k]$ centered around $k_{\rm
b}$. If the interval is not too large, i.e. $\Delta k \ll k_{\rm b}$
then, at first order in $\Delta k/k_{\rm b}$, we get
\begin{equation}
\label{rhoper}
\langle n(k) \vert \rho \vert n(k) \rangle \simeq
\frac{n}{\pi ^2}
\frac{\Delta k}{k_{\rm b}} \frac{k_{\rm b}^4}{a^4}=
\frac{n}{\pi ^2}
\frac{\Delta k}{k_{\rm b}} H_{\rm inf}^4e^{4N_{\rm e}}\, \, ,
\end{equation}
where $N_{\rm e}$ is the number of $e$-folds {\it counted back} from
the time of exit, see Fig~\ref{breac}. The time of exit is determined
by the condition $k_{\rm phys}\equiv k/a\simeq H_{\rm inf}$, where
$H_{\rm inf}$ is the Hubble parameter during inflation. It is simply
related to the scale of inflation, $M_{\rm inf}$, by the relation
$H_{\rm inf}\simeq M_{\rm inf}^2/m_{_{\rm Pl}}$. We have also assumed
that, during inflation, the scale factor behaves as $a(t)\propto
\exp(H_{\rm inf}t)$.  From Eq.~(\ref{rhoper}), we see that the
back-reaction problem occurs when one goes back in time since the
energy density of the quanta scales as $\simeq 1/a^4$. In this case,
the number of e-folds $N_{\rm e}$ increases and the quantity $\langle
n(k) \vert \rho \vert n(k)\rangle $ raises. This calculation is valid
as long as $\langle n(k) \vert \rho \vert n(k) \rangle < \rho _{\rm
inf}=m_{_{\rm Pl}}^2H_{\rm inf}^2$. When these two quantities are
equal, the energy density of the fluctuations is equal to the energy density  
of the background and the linear theory breaks down. This happens for
$N_{\rm e}=N^{\rm br}$ such that
\begin{equation}
N^{\rm br}\simeq
\frac{1}{2}\ln \biggl(\frac{m_{_{\rm Pl}}}{H_{\rm inf}}\biggr)\, ,
\end{equation}
where we have assumed $n\Delta k/(\pi ^2k_{\rm b})\simeq {\cal O}(1)$.
Interestingly enough, this number does not depend on the scale $k$ but
only on the Hubble radius during inflation $H_{\rm inf}$. This means
that for each scale considered separately, the back-reaction problem
starts to be important after the same number of e-folds $N_{\rm
e}=N^{\rm br}$ counted back from horizon exit (this is why in
Fig.~\ref{breac}, one has $N_1^{\rm br}=N_3^{\rm br}$ for the two
different scales $\lambda _1$ and $\lambda _3$). If, for instance, we
consider the case where inflation takes place at GUT scales, $M_{\rm
inf}\simeq 10^{16}\mbox{GeV}$, then $H_{\rm inf}\simeq
10^{13}\mbox{GeV}$ and one obtains $N^{\rm br}\simeq 7$ in agreement
with the estimates of Ref.~\cite{ll}. If the distribution $n(k)$ is
not strongly peaked around a particular scale but is rather spread
over a large interval, it is clear that the important mode of the
problem is, very roughly speaking, the populated smallest scale
(i.e. $\simeq \lambda _1$ in Fig.~\ref{breac}). In the following, this
scale is denoted by $\lambda _{\rm pop}$.  The value of this scale
clearly depends upon the form of the distribution $n(k)$. As it can be
seen in Fig.~\ref{breac}, $\lambda _{\rm pop}$ is the scale for which
the back-reaction problem shows up first, as we go backward in time,
since the other modes with larger wavelengths have not yet penetrated
deeply into the horizon and therefore do not yet face a back-reaction
problem. As a consequence, this scale determines the total number of
e-folds during inflation without a back-reaction problem.

Once the number $N^{\rm br}$ has been calculated, the total number of
e-folds during inflation without a back-reaction problem is {\it a
priori} fixed. It remains to be checked whether this number is still
sufficient to solve the usual problems of the hot big bang model. We
now turn to this question. Let $N_*(\lambda )$ be the number of
e-folds, for a given scale $\lambda$, between horizon exit during
inflation and the beginning of the radiation era, see
Fig.~\ref{breac}. The total number of e-folds of inflation without a
back-reaction problem is then $N_{\rm inf}\equiv N^{\rm br}+N_*$.  The
number $N_*$ is given by $N_*(\lambda)=\ln (a_0/a_*)-N_{\rm r}-N_{\rm
m}$, where $a_0$ and $a_*(\lambda )$ are the scale factor at present
time and at first horizon crossing, respectively.  The quantities
$N_{\rm r}$ and $N_{\rm m}$ are the number of e-folds during the
radiation and matter dominated epochs. The ratio $a_0/a_*$ is given by
$(\lambda /\ell _{\rm H})H_{\rm inf} /H_0$, where $\ell _{_{\rm H}}$
is the present day Hubble radius and $H_0$ is the present value of the
Hubble parameter given by $H_0/m_{_{\rm Pl}}\simeq 10^{-61}$. The
quantities $N_{\rm r}$ and $N_{\rm m}$ are given by $N_{\rm r}=\ln
(T_{\rm rh}/T_{\rm eq})$ and $N_{\rm m}\simeq \ln (z_{\rm eq})\simeq
9$.  $T_{\rm rh}$ is the reheating temperature which can be expressed
as $T_{\rm rh}\simeq (\Gamma m_{_{\rm Pl}})^{1/2}$ where $\Gamma $ is
the decay width of the inflaton. For consistency, one must have
$M_{\rm inf}\ge T_{\rm rh}$.  $T_{\rm eq}$ is the temperature at
equivalence between radiation and matter and its value reads $T_{\rm
eq}\simeq 5\times 10^{-9}\mbox{GeV} \simeq 5\times 10^{-28}m_{_{\rm
Pl}}$. The quantity $N_*(\lambda )$ can be expressed as           
\begin{equation}
N_*(\lambda )\simeq \ln \biggl(\frac{\lambda }{\ell _{_{\rm
H}}}\biggr)
+\biggl[\log _{10}\biggl(\frac{H_{\rm inf}}{m_{_{\rm Pl}}}\biggr)-
\log _{10}\biggl(\frac{T_{\rm rh}}{m_{_{\rm Pl}}}\biggr)
+29\biggr]\times \ln 10\, .
\end{equation}
From now on, in order to simplify the discussion, we assume that the
decay width of the inflaton field is such that $T_{\rm rh}\simeq
M_{\rm inf}$. Under these conditions, the usual problems are solved if
the number $N_{\rm inf}$ is such that $N_{\rm inf}(\lambda _{\rm
pop})\simeq \ln (\lambda _{\rm pop}/\ell _{_{\rm H}} )+29\times \ln
10>-4+\ln z_{\rm end}$, where the quantity $z_{\rm end}$ is the
redshift at which the standard evolution (hot big bang model)
starts. It is linked to the reheating temperature by the relation
$\log_{10}(z_{\rm end})\simeq 32+\log _{10}(T_{\rm rh}/m_{_{\rm
Pl}})$.
This gives a constraint on the
scale of inflation, namely
\begin{equation}
\label{cons}
\log _{10}\biggl(\frac{H_{\rm inf}}{m_{_{\rm Pl}}}\biggr)
< 2\log _{10}\biggl(
\frac{\lambda _{\rm pop}}{{\ell _{_{\rm H}}}}\biggr)-2.5 \, .
\end{equation}
It is known that inflation can take place between the TeV scale and
the Planck scale which amounts to $-32<\log _{10}(H_{\rm inf}/m_{_{\rm
Pl}})<0$. We see that the constraint given by Eq.~(\ref{cons}) is not
too restrictive. In particular, if we take $\lambda _{\rm pop}=0.1\ell
_{_{\rm H}}$ and $H_{\rm inf}=10^{13}\mbox{GeV}$, it is
satisfied. However, if we decrease the scale $\lambda _{\rm pop}$, the
constraint becomes more restrictive. The constraint derived in the
present article appears to be less restrictive than in Ref.~\cite{ll}
because we do not assume that all scales are populated.     

Another condition must be taken into account. We have seen that the
duration of inflation without a back-reaction problem is determined by
the evolution of $\lambda _{\rm pop}$. However, at the time at which
the back-reaction problem shows up, one must also check that all the
scales of astrophysical interest today were inside the horizon so that
physically meaningful initial conditions can be chosen. This property
is one of the most important advantages of the inflationary
scenario. If we say that the largest scale of interest today is the
horizon, this condition is equivalent to
\begin{equation}
N_*(\ell _{_{\rm H}})<N_*(\lambda _{\rm pop})+N^{\rm br} \Rightarrow
N^{\rm br}>\ln
\biggl(\frac{\ell _{_{\rm H}}}{\lambda _{\rm pop}}\biggr)\, .
\end{equation}
This condition is also not very restrictive, especially for large
scales. As previously, the condition can be more restrictive of one
wants to populate smaller scales. 

\noindent {\bf 3. Remarks:} Let us summarize now. What it is shown
above is that, roughly $N^{\rm br}$ e-folds before the relevant mode
left the horizon, we face a back-reaction problem, as the energy
density of the perturbation $\langle n(k)\vert \rho \vert n(k)
\rangle$ becomes of the same order of magnitude as the background
$\rho$.  So, before concluding that non-vacuum initial states may or
may not turn off the inflationary phase, one should calculate the
back-reaction effect, i.e., extend the present framework to second
order as it was done in Ref.~\cite{Abramo}. To our knowledge, this
analysis is still to be performed.  Moreover, even if we take the most
pessimistic position, that is, one in which we assume that the
back-reaction of the perturbations on the background energy density
prevents the inflationary phase, there still exist models of inflation
where the previous difficulties do not show up. Therefore, in the most
pessimistic situation, there is still a hope to reconcile non-vacuum
initial states with inflation. Admittedly, the price to pay is a
fine-tuning of the free parameters describing inflation and/or the
non-vacuum state.

\noindent
{\em Acknowledgments}

\noindent 
A.~G. acknowledges {\sc CONICET}, {\sc UBA} and {\sc Fundaci\'on
Antorchas} for financial support. 



\noindent {\bf References}
\begin{thebibliography}{99}
\bibitem{MRS}  J.~Martin, A.~Riazuelo and M.~Sakellariadou,
Phys.~Rev.~D {\bf 61}, 083518 (2000). 
\bibitem{ll}
A.~R.~Liddle, D.~H.~Lyth, Phys.~Reports {\bf 231}, 1 (1993),
{\tt astro-ph/9303019}.
\bibitem{Abramo}
V.~Mukhanov, L.~Abramo and R.~Brandenberger, Phys.~Rev.~Lett. {\bf 78}, 1624 (1997).
\bibitem{Unruh}
W. Unruh, {\tt astro-ph/9802323}.
\bibitem{GMS}
A.~Gangui, J.~Martin and M.~Sakellariadou, {\tt astro-ph/0205202}.
\end{thebibliography}
\end{document}